\documentclass[aps,twocolumn,prl,amssymb,superscriptaddress,floatfix]{revtex4-1}
\usepackage{amsmath}
\usepackage{graphicx}
\usepackage{color}
\usepackage{xcolor}
\usepackage{setspace}
\usepackage{multirow}
\usepackage{enumerate}
\usepackage{array}
\usepackage{multirow}
\usepackage{tabularx}
\usepackage{hyperref} \hypersetup{colorlinks=true,citecolor=blue,linkcolor=blue,urlcolor=blue}

\def\dGFAX{\chi_{\substack{\scalebox{0.6}{GFA}}}\left(\{X\}\right)}
\def\AFLOW{{\small AFLOW}}
\def\AFLOWCHULL{{\small AFLOW-CHULL}}

\def\VASP{{\small VASP}}
\def\GFA{{\small GFA}}
\def\AE{{\small AE}}
\def\GGAPBE{{\small GGA-PBE}}
\def\PAW{{\small PAW}}
\definecolor{pranab_green}{rgb}{0.31,0.53,0.10}
\definecolor{pranab_red}{rgb}{0.85,0.23,0.11}
\definecolor{PaleBlue}{rgb}{0.4,0.7,1}
\definecolor{Gray}{gray}{0.9}
\definecolor{LightCyan}{rgb}{0.88,1,1}

\makeatletter \renewcommand\frontmatter@abstractwidth{\dimexpr\textwidth\relax} \makeatother

\begin{document}

\title{\Large Metallic glasses for biodegradable implants}

\author{Denise C. Ford}
\affiliation{Department of Mechanical Engineering and Materials Science and Center for Materials Genomics, Duke University, Durham, North Carolina 27708, USA}
\author{David Hicks}
\affiliation{Department of Mechanical Engineering and Materials Science and Center for Materials Genomics, Duke University, Durham, North Carolina 27708, USA}
\author{Corey Oses}
\affiliation{Department of Mechanical Engineering and Materials Science and Center for Materials Genomics, Duke University, Durham, North Carolina 27708, USA}
\author{Cormac Toher}
\affiliation{Department of Mechanical Engineering and Materials Science and Center for Materials Genomics, Duke University, Durham, North Carolina 27708, USA}
\author{Stefano Curtarolo}
\email[]{email: stefano@duke.edu}
\affiliation{Department of Mechanical Engineering and Materials Science and Center for Materials Genomics, Duke University, Durham, North Carolina 27708, USA}
\affiliation{Departments of Electrical Engineering, Physics, and Chemistry, Duke University, Durham, North Carolina 27708, USA}

\date{\today}

\begin{abstract}
\noindent
Metallic glasses are excellent candidates for biomedical implant applications due to their inherent strength and corrosion resistance.
Use of metallic glasses in structural applications is limited, however, because bulk dimensions are challenging to achieve.
Glass-forming ability (\GFA) varies strongly with alloy composition and becomes more difficult to predict as the number of chemical species in a system increases.
Here we present a theoretical model --- implemented in the \AFLOW\ framework --- for predicting \GFA\ based on the competition between crystalline phases, and apply it to biologically relevant binary and ternary systems.
Elastic properties are estimated based on the rule of mixtures for alloy systems that are predicted to be bulk glass-formers.
Focusing on Ca- and Mg-based systems for use in biodegradable orthopedic support applications, we suggest alloys in the AgCaMg and AgMgZn families for further study; and alloys based on the compositions: Ag$_{0.33}$Mg$_{0.67}$, Cu$_{0.5}$Mg$_{0.5}$, Cu$_{0.37}$Mg$_{0.63}$ and Cu$_{0.25}$Mg$_{0.5}$Zn$_{0.25}$.
\end{abstract}
\maketitle

Metallic glasses demonstrate greater strength and corrosion resistance than their crystalline counterparts, and are therefore highly sought-after materials for a variety of applications, such as precision gears, sporting goods, and medical devices \cite{Johnson_BMG_2009, Kruzic_aem_BMGstruct_2016, Li_actabiomat_BmgAdv_2016, Kaur_jbmra_RevBioactiveGlass_2014, Inoue_imr_ironBMG_2013}.
These properties are of particular importance for biomedical implant applications --- as implant materials must maintain function and biocompatibility in a chemically and mechanically complex physiological environment.
For example, stainless steel, which is well-known for strength, corrosion resistance, and biocompatibility, is commonly used in implantable devices ({\it e.g.} vascular stents, pacemakers, and total joint replacements) \cite{Thouas_mser_metalImplant_2015}.
Materials-related failure of these devices, however, does occur and can necessitate a revision surgery.
In particular, structural support implants are plagued by fatigue and stress corrosion cracking, despite optimization of the steel by alloying and surface treatments \cite{Thouas_mser_metalImplant_2015}.
Ions are released during corrosion and structural failure, which can lead to allergic reaction, metallosis, or toxicity.

For applications where only temporary support is required, such as bone plates and coronary stents, an intentionally degradable material can be used; this reduces the occurrence of implant removal surgeries and eliminates concern over long-term embedding of a foreign object.
Mg-based alloys are considered for orthopedic applications:
densities and mechanical properties are similar to bone, and Mg is an essential nutrient for humans that is active in bone development \cite{Staiger_biomat_MgOrtho_2006}.
Despite favorable characteristics and intense study over the last two decades \cite{Walker_jbmrb_MgBioPerspec_2014, Farraro_jbm_RevolOrtho_2014, Chen_actabiomat_MgAdv_2014, Jafari_jom_MgSCC-CF_2015, Agarwal_msec_MgRevCBS_2016}, Mg-based alloys are still not in wide-spread clinical use because they often degrade too quickly forming hydrogen gas.
When the rate is too high, bubbles can interfere with tissue healing \cite{Narayanan_pms_MAO_2014, Kraus_actabiomat_Mginvivo_2012, Song_CS_BioDC_2007, Ibrahim_msec_PostFab_2017},
block the blood stream, or cause alkaline poisoning \cite{Narayanan_pms_MAO_2014, Song_CS_BioDC_2007}.

Alternative to traditional metallurgical approaches --- alloying and heat treating --- strength and corrosion resistance can be improved for a material by forming an amorphous structure.
Since glasses lack grain boundaries and dislocations, galvanic couples are reduced and slip planes are eliminated.
Ion diffusion may also be retarded by elimination of structural defects, enhancing corrosion resistance.
Many studies indicate that metallic glasses have higher yield strength than their crystalline counterparts, and the yield strength increases with the glass transition temperature \cite{Cheng_pms_SPMG_2011}.
Fe-based metallic glasses have been shown to passivate/repassivate quickly in aqueous solutions and have high resistance to pitting, potentially retarding stress corrosion cracking \cite{Scully_jmr_BMG-CMP_2007}.
No hydrogen evolution could be clinically observed for a MgZnCa metallic glass studied for biomedical implant purposes \cite{Zberg_NMAT_MgZnCa_2009}.
Elimination of structural defects, however, can be problematic for support applications because plastic deformation cannot occur.
Many metallic glasses have been shown to undergo brittle fracture \cite{Cheng_pms_SPMG_2011}, and increasing ductility is currently an active research area \cite{Lewandowski_phml_plasticMG_2005,Chen_armr_MG-duct_2008}.

Discovery of new metallic glass-forming systems has also been vigorously pursued over the past few decades.
The empirical rules devised by Inoue \cite{Inoue_stabilization_2000} have been heavily relied upon for experimental discovery.
Further guidance is needed: recent works estimate that there are $\sim$3 million potential binary, ternary, and quaternary bulk metallic glasses based on empirical rules \cite{Li_acscombsci_numMG_2017}, and that it could take up to a decade to search the ternary space of the 30 common elements with high-throughput combinatorial experimentation \cite{Ren_sciadv_MGexpML_2018}.
Physical models for glass-forming ability (\GFA) have been suggested \cite{Vincent_jncs_thermoBMG_2011, Laws_ncomm_BMGmodel_2015, zhang2015origin, LL_jap_compGFA_2010}, yet no complete, robust theory exists.
Recently, the concept of structural confusion during cooling advanced by Greer in 1993 \cite{greer1993confusion} has been explored computationally by Perim {\it et al.} \cite{curtarolo:art112} with reasonable success.
A glass-forming descriptor was devised based on similarity between the formation enthalpies of compounds available in a given system (similar probability of occurrence) and between their crystal structures (large differences creating confusion).
The descriptor was successfully applied to the \AFLOW.org repository \cite{aflowlib,curtarolo:art92,curtarolo:art75} of first-principles calculations of binary alloys.
Extension to higher-order systems, however, is not trivial.
Glass-forming ability is expected to increase with the number of species in a system because the configuration space grows \cite{LL_jap_compGFA_2010}, but this is not always the case \cite{Laws_ncomm_BMGmodel_2015, zhang2015origin}.
Since the \GFA\ of a glass directly relates to the critical dimensions of a glassy alloy, it is of great technological interest to predict materials with high \GFA\ for use in applications where bulk dimensions are required.

Here, we define and compute the \GFA\ for binary and ternary alloy systems of interest for biomedical implant applications, extending the concept of structural frustration \cite{greer1993confusion} and the approach of Perim {\it et al.} \cite{curtarolo:art112}.
In the remainder of this article, we describe the formalism for the prediction of \GFA\ and validate the approach against known bulk glass-forming systems found in the literature.
We then present predictions for glass-formation in biologically relevant alloy systems and estimate their elastic properties based on the rule of mixtures.
Finally, we recommend alloys for use in orthopedic support applications. \\ \ \\

{\noindent \bf Calculation of glass-forming ability.}
The calculation of \GFA\ is based on the formalism described in Ref. \cite{curtarolo:art112}, in which all of the metastable structures available at a given stoichiometry  vector \{$X$\} are compared to a reference state --- the structure with the lowest enthalpy of formation at \{${X}$\}.
An average structural similarity between metastable structures is also considered.
Individual structures are represented by a vector $\psi_i$ of the local atomic environments (\AE) \cite{villars:factors} calculated for each unique atom.
The \AE\ of an atom is defined as the polyhedron formed by the atoms within the maximum gap in the radial distribution function.
Functions describing the similarity between the structures $f(\left | \psi_i \right \rangle)$, similarity between the formation enthalpies of a metastable structure and the reference state $g(H_i)$, and sampling are combined as:
\begin{equation}
\dGFAX=\frac{\sum\limits_{i} f(\left | \psi_i \right \rangle) g(H_i)}{\mathrm {\#\ structures}}.
\label{GFA1}
\end{equation}
This works well as a first-order approximation for binary systems.
Here it is further developed to include contributions from local variations in stoichiometry, and the chemical identity of the central atom in each \AE.
The reference state is also more accurately described by calculating the convex hull for each system using \AFLOWCHULL~\cite{Oses_CHULL_JCIM-pub_2018}.

\begin{figure}
\centerline{\includegraphics[width=0.445\textwidth]{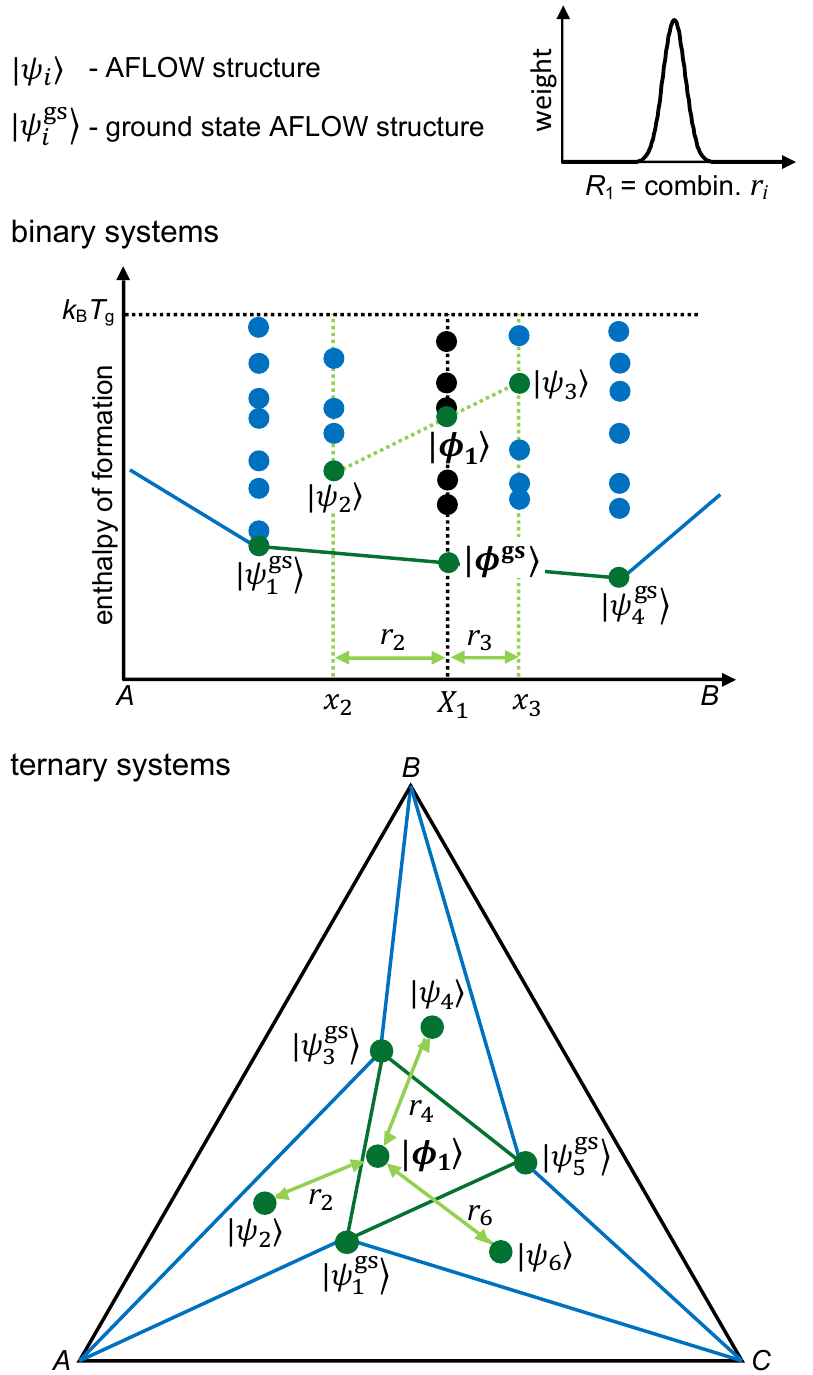}}
\caption{\small
{\bf Schematic of contributions to the \GFA.}
Structures are represented as circles: black are at the global stoichiometry, blue are at other stoichiometries, and green highlight contributing pairs and triplets.
In the ternary schematic, $\left |\phi^{\mathrm{gs}}\right \rangle$ is located directly beneath $\left |\phi_1\right \rangle$.
The cutoff energy for including structures in the analysis is based on a typical glass formation temperature ($T_{\mathrm{g}}$).
The weight of each contributing pair (triplet) is taken from a Gaussian distribution.
The reference state is defined by the convex hull.}
\label{scheme}
\end{figure}

In Perim {\it et al.} \cite{curtarolo:art112}, only the competition between structures with the same stoichiometry was considered.
In real systems, however, local variations in composition allow for structures with different stoichiometries to participate in the competition.
Here, these structures are included as linear combinations of database structures $\psi_i$ with stoichiometry $x_i$.
A combination is designated as pseudostructure $\phi_l$.
Coefficients $b_{l,i}$ are assigned to balance the local stoichiometries, $x_i$, with the global stoichiometry, $X_l$:
\begin{equation}
\sum\limits_{i} b_{l,i} x_i = X_l.
\end{equation}
The concept is depicted in Fig. \ref{scheme}.

As in Ref. \cite{curtarolo:art112}, the $\psi_i$ are comprised of sets of \AE\ calculated for each unique atom.
Each $AE_{\alpha}$ carries a coefficient, which indicates the relative amount of the \AE\ present in $\psi_i$.
Here, information about the amount of $\psi_i$ present in pseudostructure $\phi_l$ is also included in the coefficient, designated as $c_{l,\alpha}$.
Mathematically, the combination is written as:
\begin{equation}
\begin{aligned}
\left |\phi_l\right \rangle=\sum\limits_{\alpha} c_{l,\alpha}\left | AE_\alpha \right \rangle, \ \ \ \ \left \langle AE_\alpha | AE_\beta \right \rangle = \delta_{\alpha\beta}, \\
\sum\limits_{\alpha} c_{l,\alpha}^2=1, \ \ \ \ c_{l,\alpha}=\frac{\sum\limits_{i} b_{l,i} N_{i,\alpha}}{\sqrt{\sum\limits_{\alpha} (\sum\limits_{i} b_{l,i} N_{i,\alpha})^2}},
\label{pseudo}
\end{aligned}
\end{equation}
where $\delta_{\alpha\beta}$ is the Kronecker $\delta$, and $N_{i,\alpha}$ is the number of times $AE_\alpha$ occurs in $\psi_i$.
Linear combinations of pairs and triplets are created for binary and ternary alloy systems, respectively.
At \{$X$\}, the convex hull defines the ground state, which may consist of one or more structures (decomposition products).
Therefore, it is represented as a pseudostructure, $\phi^{\mathrm{gs}}$, and defined in the same way as Eqn. (\ref{pseudo}).

The similarity between two pseudostructures, $\phi_l$ and $\phi_m$, is quantified by the scalar product:
\begin{equation}
\left \langle \phi_l | \phi_m \right \rangle = \sum\limits_{\alpha\beta} \left \langle AE_\alpha \right | c_{l,\alpha}c_{m,\beta} \left | AE_\beta \right \rangle = \sum\limits_{\alpha} c_{l,\alpha}c_{m,\alpha}.
\label{scalarProduct}
\end{equation}
The chemical identity of the central atom in each \AE\ is retained, and only environments for the same chemical species are compared.
The overall comparison is a linear combination of the individual comparisons:
\begin{equation}
\left \langle \phi_l | \phi_m \right \rangle = \sum\limits_j \left \langle \phi_l | \phi_m \right \rangle_j n_j,
\end{equation}
where $n_j$ are fractions of each species at \{$X$\}.

The function which describes the structural similarity between a metastable state, $\phi_l$, and the ground state, $\phi^{\mathrm{gs}}$, is constructed to be a maximum when the pseudostructures have no {\small AE}s in common and zero when they are equivalent:
\begin{equation}
\begin{aligned}
f(\left |\phi_l\right \rangle)=|1-\left \langle \phi_l |\phi^{\mathrm{gs}}\right \rangle |w_l,
\label{morphology}
\end{aligned}
\end{equation}
where $w_l$ is the weight of each pair (triplet) contributing to the \GFA.
They follow a Gaussian distribution based on the dimensionless difference between the local and global stoichiometries, $r_i$:
\begin{equation}
r_i=|x_i-X_l|, \ \ \ \ R_l=\sum\limits_{i} r_i b_{l,i}, \ \ \ \ w_l=\exp\left(\frac{-R_l^2}{2\sigma_R^2}\right),
\label{weight}
\end{equation}
and $\sigma_R$ is a fitting factor set to 0.1.
The structural similarity amongst pairs (triplets) of metastable states contributes as a weighted average:
\begin{equation}
\begin{aligned}
\overline{ss}=\frac{\sum\limits_{lm}(1-\left \langle \phi_l |\phi_m\right \rangle)w_lw_m}{\sum\limits_{lm}}.
\end{aligned}
\end{equation}

The \GFA\ computed for each stoichiometry of a given alloy system is normalized by the sum of the weights of each contribution, $\sum\limits_{l} w_l$.

The enthalpy proximity between a metastable state and the ground state is calculated by
\begin{equation}
\begin{aligned}
g(H_l)=&\exp\left (\frac{-\left | H_l-H^{\mathrm{gs}} \right |}{k_{\mathrm B}T_{\mathrm {room}}}\right ),
\label{enthalpy}
\end{aligned}
\end{equation}
where $H_l$ is the formation enthalpy per atom of a pseudostructure, $H^{\mathrm{gs}}$ is the formation enthalpy per atom of the ground state, $k_{\mathrm{B}}$ is the Boltzmann constant, and $T_{\mathrm {room}}$ is room temperature.

Finally, the overall \GFA\ at each global stoichiometry \{$X$\} is calculated as:
\begin{equation}
\dGFAX=\frac{100 \overline{ss}^2\sum\limits_{l} f(\left | \phi_l \right \rangle) g(H_l)}{\sum\limits_{l} w_l},
\label{GFA2}
\end{equation}
where 100 is an arbitrary scaling factor. \\ \ \\

{\noindent \bf First-principles calculations of alloy prototypes.}
The crystal structures and enthalpies of formation of the alloys are taken from the \AFLOW.org repository \cite{aflowlib,curtarolo:art92,curtarolo:art75}.
The original calculations were performed systematically using the \AFLOW\ computational materials design framework \cite{aflow_fleet_chapter,aflowPAPER18,curtarolo:art65}, with the \AFLOW\ standard settings \cite{curtarolo:art104}.
These include density functional theory as implemented in \VASP\ \cite{kresse_vasp,vasp_prb1996}, the \GGAPBE\ exchange-correlation functional \cite{PBE}, \PAW\ potentials \cite{PAW,kresse_vasp_paw}, at least 6,000 k-points per reciprocal atom, and a plane-wave cutoff of at least 1.4 times the largest recommended value for the \VASP\ potentials of the constituents.
The input crystal structures were built from the \AFLOW\ library of common prototypes~\cite{aflowANRL,anrl_pt2_2018}. \\ \ \\

\begin{figure}
\centerline{\includegraphics[width=0.445\textwidth]{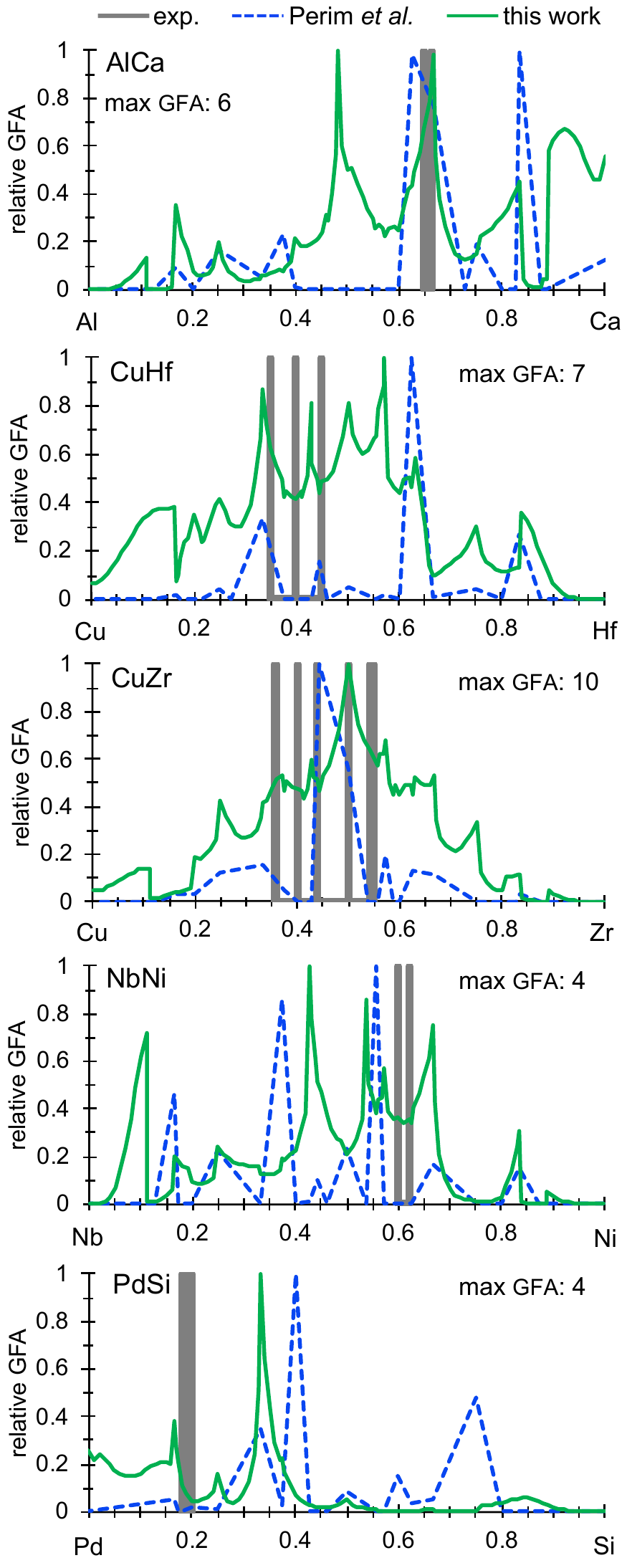}}
\caption{\small
{\bf \GFA\ predictions for known bulk binary glass-forming alloys.}
The \GFA\ predictions for AlCa, CuHf, CuZr, NbNi, and PdSi are shown as green solid lines, and compared to predictions using the method of Perim {\it et al.} \cite{curtarolo:art112} (blue dashed lines) and the \GFA\ determined experimentally (grey bars).
For the predictions, the scale is relative to the maximum \GFA\ of the system.
For the experiments, a bar indicates that a bulk alloy (critical diameter \textgreater\ 1 mm) was formed.
Experimental data was taken from \cite{Guo_apl_CaAl-BMG_2004} for AlCa, \cite{inoue2004formation,xia2006glass} for CuHf, \cite{wang2004bulk,li2008matching,inoue2004formation} for CuZr, \cite{Xia_jap_NbNi-BMG_2006,leonhardt1999solidification} for NbNi, and \cite{Chen_amet_PdSi-BMG_1969} for PdSi.}
\label{BV}
\end{figure}

\begin{figure}
\centerline{\includegraphics[width=0.445\textwidth]{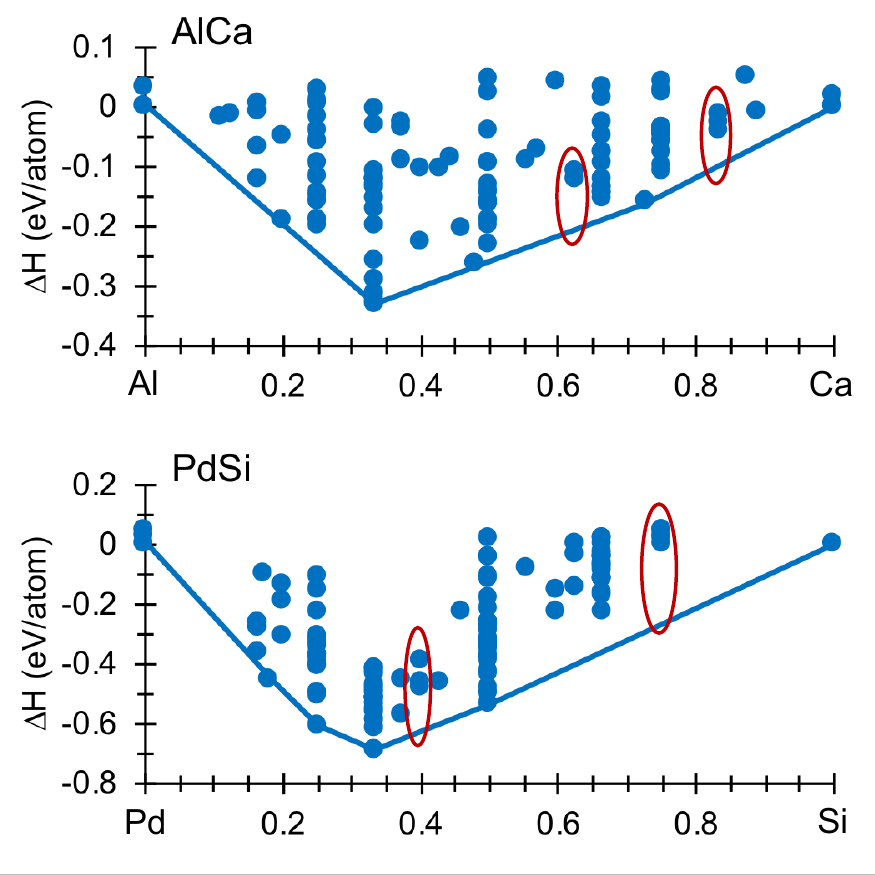}}
\caption{\small
{\bf Convex hull plots for known bulk glass-forming alloys.}
Structures and energies are from the \AFLOW.org repository \cite{aflowlib,curtarolo:art92,curtarolo:art75}.
Red ovals indicate compositions where using the convex hull to find the reference state for the \GFA\ calculations made a significant difference in the predicted \GFA.}
\label{CH}
\end{figure}

{\noindent \bf Model validation.}
The \GFA\ predictions made with the current model are shown in Fig. \ref{BV} for five binary alloy systems known to form bulk metallic glasses.
The compositions where glasses have been produced with a critical diameter of greater than 1 mm \cite{Guo_apl_CaAl-BMG_2004, inoue2004formation, xia2006glass, wang2004bulk, li2008matching, Xia_jap_NbNi-BMG_2006, leonhardt1999solidification, Chen_amet_PdSi-BMG_1969} are also marked.
In all cases, \GFA\ greater than 1/3 of the system maximum is predicted at or near the experimentally determined bulk glass-forming compositions.
The predictions made using the method of Perim {\it et al.} \cite{curtarolo:art112} are shown for comparison, where peaks are often found at or near the experimental compositions.
In some instances, there is a significant difference between the \GFA\ predicted using the two computational methods which can be attributed to the change in reference state.
These are highlighted in Fig. \ref{CH} for AlCa and PdSi.
At each highlighted composition a large peak is predicted by the method in Ref. \cite{curtarolo:art112}, which uses the structure with the lowest enthalpy of formation at a given stoichiometry for the reference state.
Because the convex hull calculation indicates a phase separation at these compositions, using it to find the reference state reduces the predicted \GFA.
Other improvements to the predictions can be attributed to including contributions from local off-stoichiometry clusters, and to a lesser extent, changes in the comparison of the {\small AE}s.
The precision of the calculation is also increased by using the current model, since the \GFA\ can be calculated at compositions where no structures are available in the database.

\begin{figure*}
\centerline{\includegraphics[width=1\textwidth]{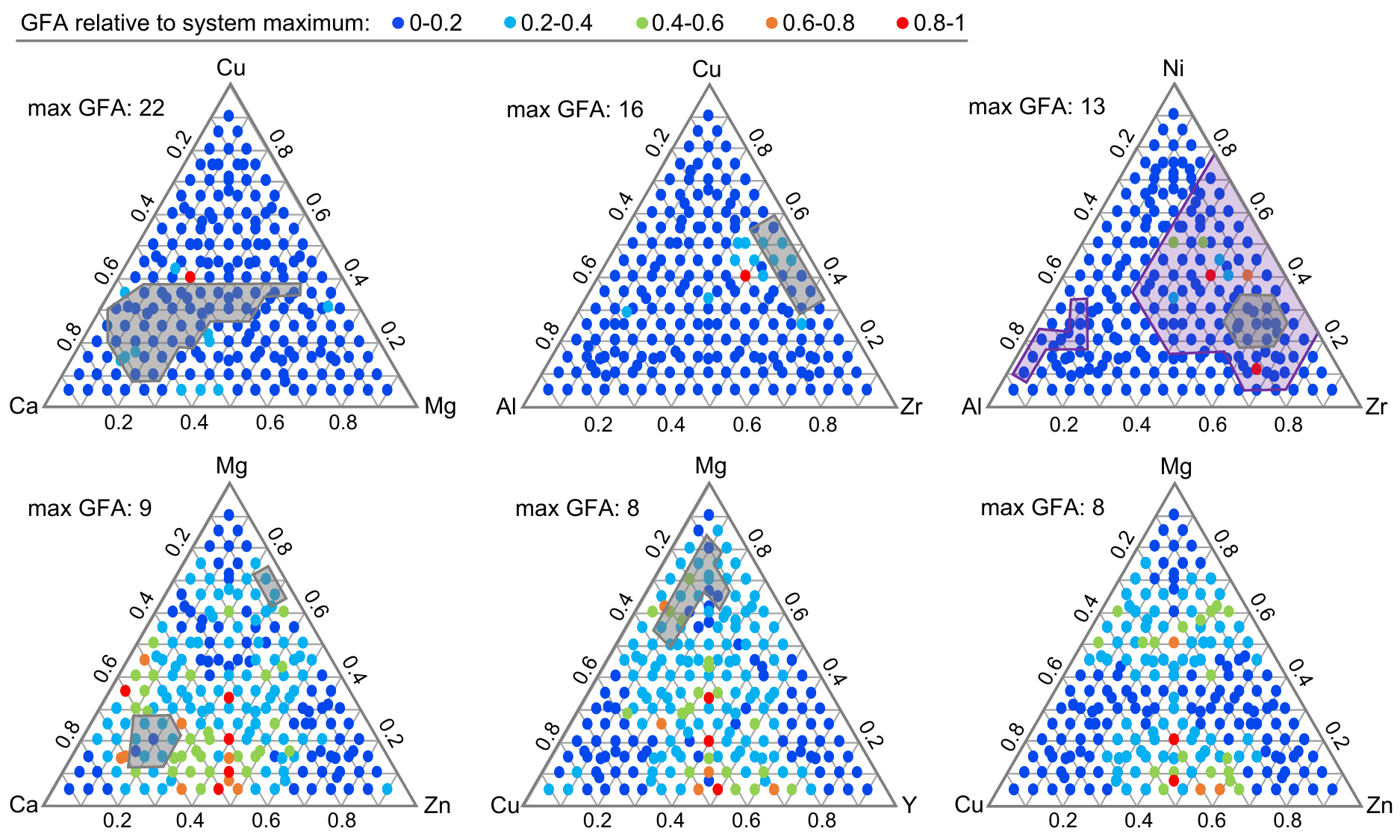}}
\caption{
{\bf \GFA\ predictions for bulk ternary glass-forming alloys.}  The \GFA\ predictions for CaCuMg, AlCuZr, AlNiZr, CaMgZn, CuMgY, and CuMgZn systems are shown as circles and approximate regions where bulk glasses have been produced are enclosed in grey shapes \cite{Wang_actamat_AlCuZr_2005,Kim_msea_AlCuZr-1_2006,Xu_prl_AlCuZr-2_2004,Inoue_mattrans_AlCuZr-3_1995,Li_jac_AlNiZr-2_2012,Senkov_jac_CaMgCu-BMG_2006,Laws_jac_AgCaCuMg_2012,Gu_jmr_MgCaZn_2005,Senkov_jncs_CaMgZn-BMG_2005,Park_jmr_CaMgZn-2_2004,Ma_jmr_CuMgY-1_2005,Inoue_mattrans_MgCuY_1991}.
 The purple shapes in the AlNiZr plot indicate regions of potential bulk glass-formation based on evaluation of the supercooled liquid \cite{Inoue_mattrans_AlNiZr-1_1990}.}
\label{TV}
\end{figure*}

\GFA\ predictions for the known bulk glass-forming systems --- AlCuZr \cite{Kim_msea_AlCuZr-1_2006,Xu_prl_AlCuZr-2_2004,Inoue_mattrans_AlCuZr-3_1995,Wang_actamat_AlCuZr_2005}, AlNiZr \cite{Inoue_mattrans_AlNiZr-1_1990,Li_jac_AlNiZr-2_2012}, CaCuMg \cite{Senkov_jac_CaMgCu-BMG_2006,Laws_jac_AgCaCuMg_2012}, CaMgZn \cite{Gu_jmr_MgCaZn_2005,Senkov_jncs_CaMgZn-BMG_2005,Park_jmr_CaMgZn-2_2004}, and CuMgY \cite{Ma_jmr_CuMgY-1_2005,Inoue_mattrans_MgCuY_1991} --- are shown in Fig. \ref{TV}.
To reduce calculation time, a cutoff of $r_i=3\sigma_R$ is imposed for including a pseudostructure in the \GFA\ calculation.
Testing was performed for the CaCuMg, CaMgZn, and CuMgY systems, which indicates an insignificant effect on the predictions.
A maximum \GFA\ of at least 4 is predicted for all five systems where bulk alloys have been produced, consistent with the results for binary alloy systems (Fig. \ref{BV}).
Regions where bulk glasses have been produced tend to be near compositions where high \GFA\ is predicted, and additional compositions may be possible (Fig. \ref{TV}).
Experimentally, regions of potential bulk-glass formation in the AlNiZr system have been predicted by evaluating the supercooled liquid \cite{Inoue_mattrans_AlNiZr-1_1990} and are outlined in purple.
Reports emphasize that \GFA\ changes rapidly with composition \cite{Ma_jmr_CuMgY-1_2005}, and as little as 1\% change in composition could change a result from amorphous to crystalline \cite{Wang_actamat_AlCuZr_2005}.
Therefore, it is important to comprehensively search the phase space.
A broad search is performed computationally, and high \GFA\ compositions far from experimentally confirmed glasses are predicted.
These compositions offer a guide for experimentalists, but are limited in precision due to the sampling grid and available crystal structures. \\ \ \\

\begin{figure*}
\centerline{\includegraphics[width=1\textwidth]{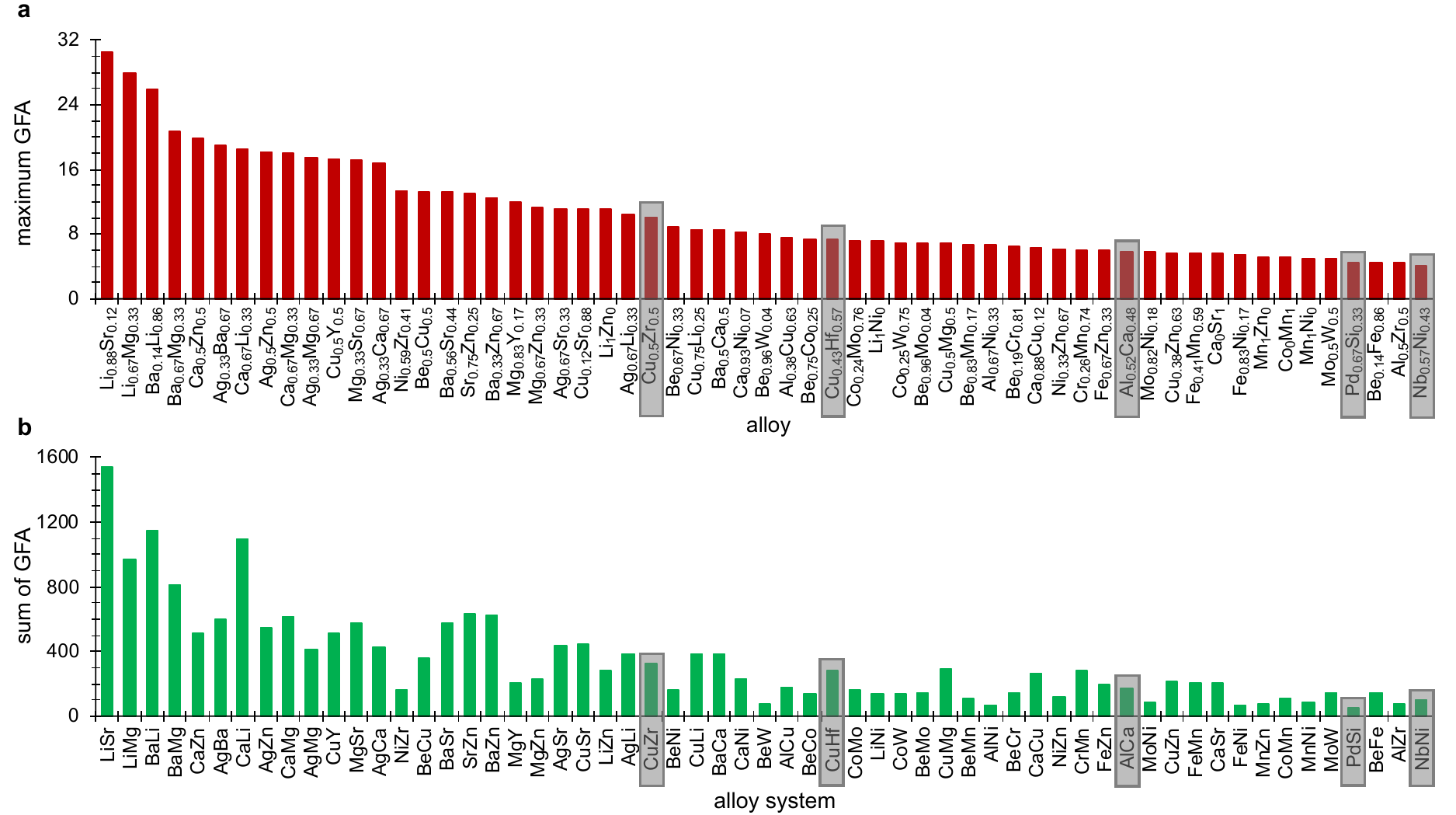}}
\caption{
{\bf \GFA\ predictions for biologically relevant binary alloy systems.}
The system maximum ({\bf a}) and sum of \GFA\ ({\bf b}).
 The known bulk glass-forming systems (enclosed in grey rectangles) and systems relevant to the ternary validation systems are shown for comparison.}
\label{BP}
\end{figure*}

{\noindent \bf Calculation of elastic properties.}
The elastic properties of metallic glasses have been extensively reviewed and can be estimated from the rule of mixtures \cite{Wang_pms_elasticMG_2012,Zhang_jac_elastic-correlations_2007}.
Here the following equation is used:
\begin{equation}
M^{-1}=\sum\limits_{j} f_j M_j^{-1},
\label{mixing}
\end{equation}
where $M$ is the elastic constant of the system, and $f_j$ and $M_j$ are the molar fraction and elastic constant of the component elements in bulk form.
The elastic properties for the elements are taken from Ref. \cite{webelements}.
This method generally under-predicts the bulk modulus and over-predicts the shear modulus, $K$ and $G$, respectively.
Performing a linear regression on data for the 18 binary and ternary systems available in Ref. \cite{Wang_pms_elasticMG_2012} which are relevant to the current study gives the equations:
\begin{equation}
K_{\mathrm{measured}}=1.12K_{\mathrm{mixing}}-4.68, \ \ \ \ R^2=0.97
\label{bulk}
\end{equation}
\begin{equation}
G_{\mathrm{measured}}=0.80G_{\mathrm{mixing}}+2.92, \ \ \ \ R^2=0.93
\label{shear}
\end{equation}
Poisson's ratio ($\nu$) can be used to assess brittle {\it vs.} ductile behavior, as metallic glasses with $\nu < 0.34$ tend to exhibit brittle behavior \cite{Wang_pms_elasticMG_2012,Zhang_jac_elastic-correlations_2007}.
Since metallic glasses are macroscopically isotropic, Poisson's ratio is calculated as:
\begin{equation}
\nu_{\mathrm{calc}}=\frac{3K-2G}{2(3K+G)}
\label{nu_calc}
\end{equation}
Using the $K_{\mathrm{measured}}$ and $G_{\mathrm{measured}}$ calculated from Eqns. (\ref{bulk}) and (\ref{shear}) in Eqn. (\ref{nu_calc}) correlates well with the measured Poisson's ratios reported in Ref. \cite{Wang_pms_elasticMG_2012}:
\begin{equation}
\nu_{\mathrm{measured}}=1.10\nu_{\mathrm{calc}}-0.04, \ \ \ \ R^2=0.80
\label{nu}
\end{equation}
Another commonly applied indicator of brittle {\it vs.} ductile behavior, the Pugh's modulus ratio ($G$/$K$), is less reliably predicted from Eqns. (\ref {mixing}-\ref {shear}).
The Young's modulus ($E$) correlates with fracture strength ($\sigma_{\mathrm{f}}$) as $\sigma_{\mathrm{f}}$=$E/50$, and $E$ is calculated as $E=2.61G$ \cite{Wang_pms_elasticMG_2012}.  \\ \ \\

{\noindent \bf Application to biologically relevant systems.}
The \GFA\ for binary combinations of macronutrient metals (Ca, Mg), trace metals in the human body (Ba, Be, Co, Cr, Cu, Fe, Li, Mn, Mo, Ni, Sr, Zn, W), and antibacterial Ag \cite{Thouas_mser_metalImplant_2015,Vasconcelos_biomat_twoface_2016,Kaur_jbmra_RevBioactiveGlass_2014} is calculated.
Sampling for these binary systems ranges from 134 to 910 converged crystal structures.
Upon removal of duplicates (one converged structure is equivalent to another converged structure \cite{aflowsym,aflow_compare-updated_2018}), structures with formation enthalpy greater than 0.05 eV ($\sim$600 K, a typical critical temperature for glass formation), and structures for which {\small AE}s could not be determined; 4 to 567 structures remain for each system for calculation of the \GFA.
Finally, systems with less than 20 remaining structures are considered to have insufficient sampling and removed from the analysis.
This leaves 55 systems, of which 49 have a maximum \GFA\ \textgreater\ 4 (the smallest maximum for known binary bulk glass-forming systems) and 21 have a maximum \GFA\ \textgreater\ 10 (the largest maximum for the known binary bulk glass-forming systems).
The maximum \GFA\ and the composition at which it occurs for these systems is shown in Fig. \ref{BP}.
Metallic glasses have been produced experimentally for ten of them --- AgCa \cite{Amand_smet_CaSr-based-MG_1978}, CaCu \cite{Amand_smet_CaSr-based-MG_1978}, CaMg \cite{Amand_smet_CaSr-based-MG_1978}, CuMg \cite{Kim_mattrans_CuMg-MG_1990}, CaZn \cite{Amand_smet_CaSr-based-MG_1978}, MgSr \cite{Amand_smet_CaSr-based-MG_1978}, MgZn \cite{Calka_smet_MgZn-MG_1977}, and SrZn \cite{Amand_smet_CaSr-based-MG_1978}.

The ternary system CuMgZn is also studied.
Sampling for the ternary systems ranges from 1276 to 1399 converged crystal structures.
Upon removal of duplicates \cite{aflowsym,aflow_compare-updated_2018}, structures with formation enthalpy greater than 0.05 eV, and structures for which {\small AE}s could not be determined; 399 to 580 structures remain for each system for calculation of the \GFA.
The predictions are shown in Fig. \ref{TV}.
CuMgZn has a maximum predicted \GFA\ of 8, therefore is expected to be a bulk glass-forming system.

\begin{table*}
\caption{Ca and Mg rich peaks in biologically relevant binary systems with a predicted \GFA\ at least 1/3 of the system maximum.}
\begin{tabular}{ll|ll}
\hline
Alloy \ \ \ \ & Relative \GFA\ \ \ \ \ \ \ \ \ & Alloy \ \ \ \  & Relative \GFA\ \\
\hline
Ag$_{0.33}$Ca$_{0.67}$ \ \ \ \ & 1 \ \ \ \ \ \ \ \ & Ca$_{0.57}$Mg$_{0.43}$ \ \ \ \ & 0.59 \\
Ag$_{n}$Ca$_{1-n}$, $0.05<n<0.14$ \ \ \ \ & 0.33-0.38 \ \ \ \ \ \ \ \ & Ca$_{0.5}$Mg$_{0.5}$ \ \ \ \ & 0.73 \\
Ag$_{0.37}$Mg$_{0.63}$ \ \ \ \ & 0.74 \ \ \ \ \ \ \ \ & Ca$_{1}$Sr$_{0}$ \ \ \ \ & 0.58 \\
Ag$_{0.33}$Mg$_{0.67}$ \ \ \ \ & 1 \ \ \ \ \ \ \ \ & Ca$_{0.75}$Sr$_{0.25}$ \ \ \ \ & 0.62 \\
Ag$_{0.25}$Mg$_{0.75}$ \ \ \ \ & 0.46 \ \ \ \ \ \ \ \ & Ca$_{0.67}$Sr$_{0.33}$ \ \ \ \ & 0.44 \\
Ag$_{0.2}$Mg$_{0.8}$ \ \ \ \ & 0.38 \ \ \ \ \ \ \ \ & Ca$_{0.57}$Sr$_{0.43}$ \ \ \ \ & 0.72 \\
Ag$_{n}$Mg$_{1-n}$, $0.05<n<0.12$ \ \ \ \ & 0.34-0.4 \ \ \ \ \ \ \ \ & Ca$_{0.5}$Sr$_{0.5}$ \ \ \ \ & 0.72 \\
Ba$_{0.5}$Ca$_{0.5}$ \ \ \ \ & 1 \ \ \ \ \ \ \ \ \ & Ca$_{0.65}$Zn$_{0.35}$ \ \ \ \ & 0.7 \\
Ba$_{0.44}$Ca$_{0.56}$ \ \ \ \ & 0.67 \ \ \ \ \ \ \ \ & Ca$_{0.5}$Zn$_{0.5}$ \ \ \ \ & 1 \\
Ba$_{0.33}$Ca$_{0.67}$ \ \ \ \ & 0.68 \ \ \ \ \ \ \ \ & Cu$_{0.5}$Mg$_{0.5}$ \ \ \ \ & 1 \\
Ba$_{0.5}$Mg$_{0.5}$ \ \ \ \ & 0.71 \ \ \ \ \ \ \ \ & Cu$_{0.37}$Mg$_{0.63}$ \ \ \ \ & 0.93 \\
Ba$_{0.17}$Mg$_{0.83}$ \ \ \ \ & 0.49 \ \ \ \ \ \ \ \ & Cu$_{0.33}$Mg$_{0.67}$ \ \ \ \ & 0.89 \\
Ca$_{0.88}$Cu$_{0.12}$ \ \ \ \ & 1 \ \ \ \ \ \ \ \ & Cu$_{0.25}$Mg$_{0.75}$ \ \ \ \ & 0.47 \\
Ca$_{0.76}$Cu$_{0.24}$ \ \ \ \ & 0.58 \ \ \ \ \ \ \ \ & Cu$_{n}$Mg$_{1-n}$, $0<n<0.16$ \ \ \ \ & 0.4-0.54 \\
Ca$_{0.67}$Cu$_{0.33}$ \ \ \ \ & 0.6 \ \ \ \ \ \ \ \  & Li$_{n}$Mg$_{1-n}$, $0.33<n<0.5$ \ \ \ \ & 0.41-0.55 \\
Ca$_{0.62}$Cu$_{0.38}$ \ \ \ \ & 0.69 \ \ \ \ \ \ \ \ & Mg$_{n}$Sr$_{1-n}$, $0.9<n<0.97$ \ \ \ \ & 0.36-0.42 \\
Ca$_{n}$Li$_{1-n}$, $0.84<n<0.9$ \ \ \ \ & 0.43-0.51 \ \ \ \ \ \ \ \ & Mg$_{0.83}$Sr$_{0.17}$ \ \ \ \ & 0.55 \\
Ca$_{0.75}$Li$_{0.25}$ \ \ \ \ & 0.9 \ \ \ \ \ \ \ \ & Mg$_{0.5}$Sr$_{0.5}$ \ \ \ \ & 0.66 \\
Ca$_{0.67}$Li$_{0.33}$ \ \ \ \ & 1 \ \ \ \ \ \ \ \ & Mg$_{0.75}$Zn$_{0.25}$ \ \ \ \ & 0.4 \\
Ca$_{0.5}$Li$_{0.5}$ \ \ \ \ & 0.84 \ \ \ \ \ \ \ \ & Mg$_{0.67}$Zn$_{0.33}$ \ \ \ \ & 1 \\
Ca$_{0.75}$Mg$_{0.25}$ \ \ \ \ & 0.35 \ \ \ \ \ \ \ \ & Mg$_{0.5}$Zn$_{0.5}$ \ \ \ \ & 0.4 \\
Ca$_{0.67}$Mg$_{0.33}$ \ \ \ \ & 1 \\
\hline
\end{tabular}
\label{peaks}
\end{table*}

\begin{figure}
\centering
\includegraphics[width=0.445\textwidth]{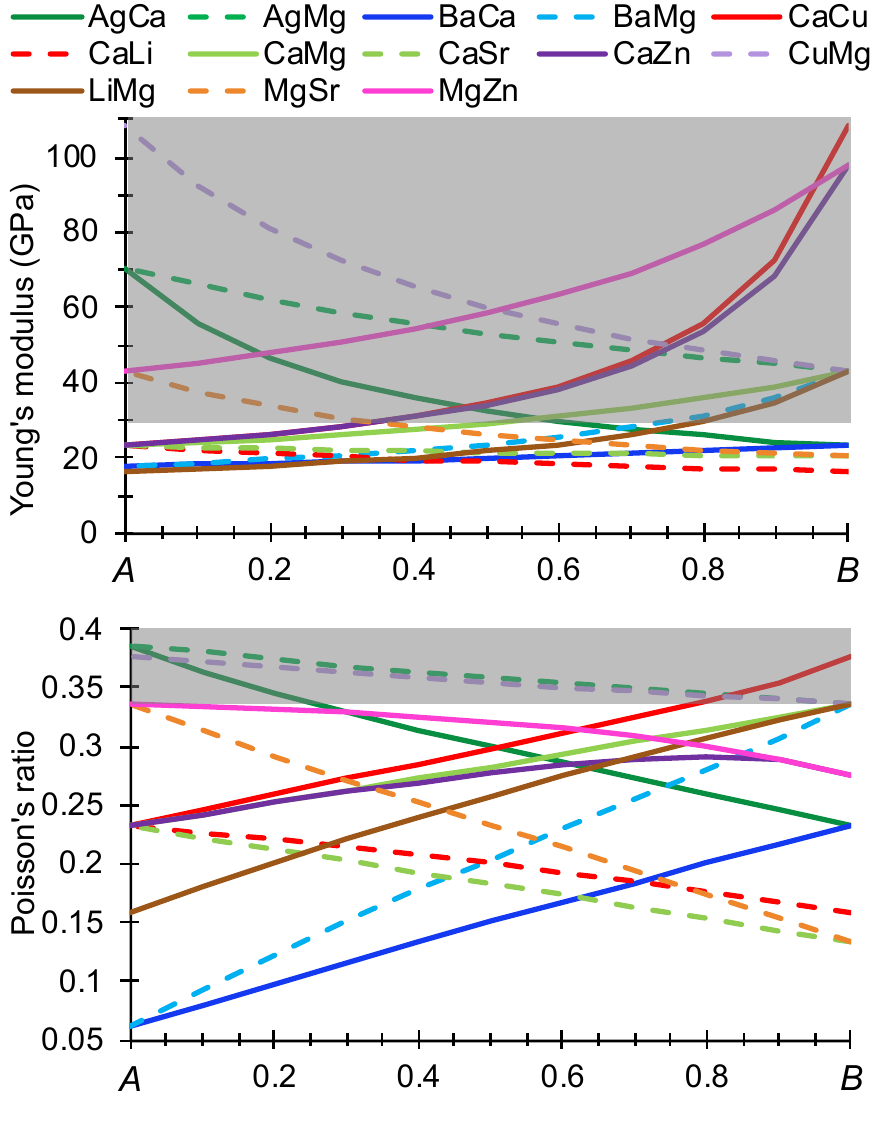}
\caption{\small
{\bf Elastic properties of biologically relevant binary alloy systems.}
The Young's moduli and Poisson's ratios for {\it AB} binary alloy systems are determined from the linear regression of experimental data with the rule of mixtures.
See text for details.
The grey regions indicate Young's moduli greater than bone \cite{Jafari_jom_MgSCC-CF_2015} and the expected region of ductility.}
\label{moduli}
\end{figure}

Because macronutrient Mg and Ca are known to be good glass-formers \cite{Senkov_jac_CaMgCu-BMG_2006,Laws_jac_AgCaCuMg_2012,Senkov_jncs_CaMgZn-BMG_2005,Amand_smet_CaSr-based-MG_1978,Calka_smet_MgZn-MG_1977,Park_jmr_CaMgZn-2_2004,Ma_jmr_CuMgY-1_2005,Inoue_mattrans_MgCuY_1991,Kim_mattrans_CuMg-MG_1990,Gu_jmr_MgCaZn_2005} and these elements are used by the human body for bone growth and repair \cite{Kaur_jbmra_RevBioactiveGlass_2014}, we focus the rest of this discussion on Mg- and Ca- based alloys for orthopedic applications.
\GFA\ peaks greater than 1/3 of the system maximum for Mg-rich and Ca-rich binary alloys are presented in Table \ref{peaks}.
The Young's moduli and Poisson's ratios for these alloy systems are plotted in Fig. \ref{moduli}, and the Young's moduli and Poisson's ratios for the best glass-forming compositions of the relevant ternary alloy systems --- CaCuMg, CaMgZn, and CuMgZn --- are given in Table \ref{Tprop}.

\begin{table}
\caption{\small
\GFA\, Young's modulus ($E$), and Poisson's ratio ($\nu$) for CaCuMg, CaMgZn, and CuMgZn compositions where the \GFA\ is greater than 60\% of the system maximum.
The Young's moduli and Poisson's ratios are determined from the linear regression of experimental data with the rule of mixtures.
See text for details.}
\centering
\begin{tabular}{lccc}
\hline
Alloy   \ \ \ \ \ \ \ \ \ \                              & \GFA\  \ \ \ \ \ \   & $E$ (GPa)  \ \ \ \ \ \   & $\nu$ \\
\hline
Ca$_{0.4}$Cu$_{0.4}$Mg$_{0.2}$ \ \ \ \ \ \ \ \ \ \       & 22 \ \ \ \ \       & 36   \ \ \ \ \ \       & 0.31 \\
\hline
Ca$_{0.33}$Mg$_{0.33}$Zn$_{0.33}$ \ \ \ \ \ \ \ \ \ \    & 9 \ \ \ \ \ \      & 37   \ \ \ \ \ \       & 0.30 \\
Ca$_{0.5}$Mg$_{0.05}$Zn$_{0.45}$ \ \ \ \ \ \ \ \ \ \     & 9 \ \ \ \ \ \      & 33   \ \ \ \ \ \       & 0.28 \\
Ca$_{0.4}$Mg$_{0.2}$Zn$_{0.4}$ \ \ \ \ \ \ \ \ \ \       & 9 \ \ \ \ \ \      & 36   \ \ \ \ \ \       & 0.29 \\
Ca$_{0.45}$Mg$_{0.1}$Zn$_{0.45}$ \ \ \ \ \ \ \ \ \ \     & 8 \ \ \ \ \ \      & 35   \ \ \ \ \ \       & 0.28 \\
Ca$_{0.6}$Mg$_{0.35}$Zn$_{0.05}$ \ \ \ \ \ \ \ \ \ \     & 8 \ \ \ \ \ \      & 28   \ \ \ \ \ \       & 0.27 \\
Ca$_{0.6}$Mg$_{0.05}$Zn$_{0.35}$ \ \ \ \ \ \ \ \ \ \     & 8 \ \ \ \ \ \      & 30   \ \ \ \ \ \       & 0.27 \\
Ca$_{0.45}$Mg$_{0.05}$Zn$_{0.5}$ \ \ \ \ \ \ \ \ \ \     & 7 \ \ \ \ \ \      & 35   \ \ \ \ \ \       & 0.28 \\
Ca$_{0.71}$Mg$_{0.14}$Zn$_{0.14}$ \ \ \ \ \ \ \ \ \ \    & 7 \ \ \ \ \ \      & 27   \ \ \ \ \ \       & 0.26 \\
Ca$_{0.46}$Mg$_{0.08}$Zn$_{0.46}$ \ \ \ \ \ \ \ \ \ \    & 6 \ \ \ \ \ \      & 35   \ \ \ \ \ \       & 0.28 \\
Ca$_{0.7}$Mg$_{0.15}$Zn$_{0.15}$ \ \ \ \ \ \ \ \ \ \     & 6 \ \ \ \ \ \      & 27   \ \ \ \ \ \       & 0.26 \\
Ca$_{0.43}$Mg$_{0.14}$Zn$_{0.43}$ \ \ \ \ \ \ \ \ \ \    & 6 \ \ \ \ \ \      & 35   \ \ \ \ \ \       & 0.28 \\
Ca$_{0.5}$Mg$_{0.45}$Zn$_{0.05}$ \ \ \ \ \ \ \ \ \ \     & 6 \ \ \ \ \ \      & 30   \ \ \ \ \ \       & 0.28 \\
Ca$_{0.5}$Mg$_{0.25}$Zn$_{0.25}$ \ \ \ \ \ \ \ \ \ \     & 6 \ \ \ \ \ \      & 31   \ \ \ \ \ \       & 0.28 \\
\hline
Cu$_{0.46}$Mg$_{0.08}$Zn$_{0.46}$ \ \ \ \ \ \ \ \ \ \    & 8 \ \ \ \ \ \      & 92   \ \ \ \ \ \       & 0.33 \\
Cu$_{0.4}$Mg$_{0.2}$Zn$_{0.4}$ \ \ \ \ \ \ \ \ \ \       & 6 \ \ \ \ \ \      & 79   \ \ \ \ \ \       & 0.33 \\
Cu$_{0.35}$Mg$_{0.05}$Zn$_{0.6}$ \ \ \ \ \ \ \ \ \ \    & 6 \ \ \ \ \ \      & 94   \ \ \ \ \ \       & 0.31 \\
Cu$_{0.4}$Mg$_{0.05}$Zn$_{0.55}$ \ \ \ \ \ \ \ \ \ \     & 5 \ \ \ \ \ \      & 95   \ \ \ \ \ \       & 0.32 \\
Cu$_{0.25}$Mg$_{0.5}$Zn$_{0.25}$ \ \ \ \ \ \ \ \ \ \     & 5 \ \ \ \ \ \      & 59   \ \ \ \ \ \       & 0.34 \\
\hline
\end{tabular}
\label{Tprop}
\end{table}

Many of the glasses presented in Tables \ref{peaks} and \ref{Tprop} will be strong enough for orthopedic applications, as the Young's modulus of bone ranges from 3 to 30 GPa \cite{Jafari_jom_MgSCC-CF_2015}.
Most of these, however, are predicted to be brittle.
This is not surprising, considering that the Poisson's ratio of pure elemental Ca is 0.31 and Mg is 0.29 \cite{webelements}; and many of the Ca- and Mg- based metallic glasses studied experimentally exhibit brittle behavior \cite{Guo_scrmat_Mg-duct_2007,Widom_prb_CaMG-elastic_2011,Laws_actamat_MgDuct_2016}.
Moreover, some glasses have shown bending ductility in ribbon form after production, but embrittle at room-temperature after only a few days or weeks \cite{Calka_smet_MgZn-MG_1977,Laws_actamat_MgDuct_2016}.
Researchers have suggested alloying with elements that minimize charge transfer \cite{Widom_prb_CaMG-elastic_2011,Laws_actamat_MgDuct_2016} to increase ductility and elements with higher melting temperatures to increase stability \cite{Laws_actamat_MgDuct_2016}.
We suggest alloys with high predicted Poisson's ratios.

Following the analyses of \GFA\ and elastic properties, Ag$_{0.37}$Mg$_{0.63}$, Ag$_{0.33}$Mg$_{0.67}$, Cu$_{0.5}$Mg$_{0.5}$, Cu$_{0.37}$Mg$_{0.63}$, Cu$_{0.33}$Mg$_{0.67}$, and Cu$_{0.25}$Mg$_{0.5}$Zn$_{0.25}$ alloys are predicted to have high \GFA, strength, and ductility.
Therefore these alloys are recommended for further consideration as biomedical implant materials for orthopedic applications.
Additional ternary systems are suggested based on the results of the binary system \GFA\ calculation and estimation of elastic properties.

The sum of the predicted \GFA\ values for each of the six ternary systems included in this study is found to correlate linearly with the sum of the \GFA\ in their component binary systems.
Since the sum of the \GFA\ for a system comprises the aspects of maximum and breadth, it can be used as a rough indicator of potential good higher-order glass-forming systems.
Considering ternary systems which can be created from the set \{Ag, Ba, Be, Ca, Co, Cr, Cu, Fe, Li, Mg, Mn, Mo, Ni, Sr, Zn, W\}, include Ca and/or Mg, and include only binary pairs with sufficient sampling and \GFA\ \textgreater\ 4; there are 31 additional potential glass-forming systems available for further analysis.
Including the six validation systems, the rank of \GFA\ sum from best to worst is \{LiMgSr, BaLiMg, CaLiSr, CaLiMg, BaCaLi, BaMgSr, AgCaLi, CaLiZn, AgBaMg, BaCaMg, AgLiMg, CaCuLi, BaMgZn, CuLiMg, BaCaZn, AgCaZn, LiMgZn, CaLiNi, AgCaMg, MgSrZn, AgMgSr, AgBaCa, CaMgSr, CaMgZn, CaSrZn, CuMgSr, AgMgZn, CaCuMg, BaCaSr, AgCaSr, CuMgY, CaCuZn, CaCuSr, CaNiZn, CuMgZn, AlCuZr, AlNiZr\}.
The \GFA\ sums for their component binary systems are given in Fig. \ref{BP}.

Estimations of Poisson's ratios indicate that mostly Ag-rich alloys will fall into the ductile regime, although some Mg-rich alloys in the AgCaMg and AgMgZn
systems have an estimated Poisson's ratio greater than 0.33 and Young's modulus greater than 40 GPa.
 Some experimental work has been performed for Ag addition to Ca- and Mg- based alloys.
 Partial substitution of Ag for Cu in Cu$_{0.25}$Mg$_{0.65}$Y$_{0.1}$ increased the \GFA\ \cite{Park_jncs_AgCuMgY_2001}, and substitution of 1-3\% Ag for Zn in Ca$_{0.04}$Mg$_{0.66}$Zn$_{0.3}$ decreased the \GFA\ but increased the corrosion resistance \cite{Li_jncs_AgCaMgZn_2015}.
 A wide composition range of AgCaMg bulk metallic glasses have been produced \cite{Amiya_mattrans_AgCaCuMg_2002,Laws_jac_AgCaCuMg_2012}.
 Based on these promising theoretical and experimental results, we suggest further exploration of AgCaMg and AgMgZn alloy systems. \\  \ \\

{\noindent \bf Conclusions.}
A model to predict the glass-forming ability (\GFA) of binary and ternary alloys systems was developed based on the competition between crystalline phases, and implemented in the \AFLOW\ framework.
The model was applied to material systems relevant for biodegradable orthopedic support implants.
Alloys predicted to have high \GFA\ were further analyzed for elastic properties based on the rule of mixtures.
Alloys predicted to have high \GFA, a Young's modulus at least as high as bone, and be ductile based on the Poisson's ratio include: Ag$_{0.33}$Mg$_{0.67}$, Cu$_{0.5}$Mg$_{0.5}$, Cu$_{0.37}$Mg$_{0.63}$ and Cu$_{0.25}$Mg$_{0.5}$Zn$_{0.25}$.
Finally, the \GFA\ of binary systems was correlated with the \GFA\ of ternary systems.
Based on this analysis and the analysis of elastic properties, the AgCaMg and AgMgZn alloy systems are recommended for further study.  \\  \ \\

{\noindent \bf Data availability.}
All of the {\it ab-initio} alloy data is freely available to the public as part of the \AFLOW\ online repository and can be accessed through {\sf www.aflow.org}
following the REST-API interface \cite{curtarolo:art92}, and using the {\small AFLUX} search language \cite{curtarolo:art128}. \\ \ \\

\newcommand{\Ozolins}{Ozoli\c{n}\v{s}}


\begin{thebibliography}{10}
\expandafter\ifx\csname urlstyle\endcsname\relax
  \providecommand{\doi}[1]{doi:\discretionary{}{}{}#1}\else
  \providecommand{\doi}{doi:\discretionary{}{}{}\begingroup
  \urlstyle{rm}\Url}\fi
\providecommand{\selectlanguage}[1]{\relax}
\providecommand{\bibAnnoteFile}[1]{%
  \IfFileExists{#1}{\begin{quotation}\noindent\textsc{Key:} #1\\
  \textsc{Annotation:}\ \input{#1}\end{quotation}}{}}
\providecommand{\bibAnnote}[2]{%
  \begin{quotation}\noindent\textsc{Key:} #1\\
  \textsc{Annotation:}\ #2\end{quotation}}

\bibitem{Johnson_BMG_2009}
W.~L. Johnson, \emph{Bulk Glass-Forming Metallic Alloys: Science and
  Technology}, MRS\ Bull. \textbf{24}, 42--56 (1999).
\input{Johnson_BMG_2009}

\bibitem{Kruzic_aem_BMGstruct_2016}
J.~J. Kruzic, \emph{Bulk Metallic Glasses as Structural Materials: A Review},
  Adv.\ Eng.\ Mater. \textbf{18}, 1308--1331 (2016).
\input{Kruzic_aem_BMGstruct_2016}

\bibitem{Li_actabiomat_BmgAdv_2016}
H.~F. Li and Y.~F. Zheng, \emph{Recent advances in bulk metallic glasses for
  biomedical applications}, Acta\ Biomater. \textbf{36}, 1--20 (2016).
\input{Li_actabiomat_BmgAdv_2016}

\bibitem{Kaur_jbmra_RevBioactiveGlass_2014}
G.~Kaur, O.~P. Pandey, K.~Singh, D.~Homa, B.~Scott, and G.~Pickrell, \emph{A
  review of bioactive glasses: Their structure, properties, fabrication, and
  apatite formation}, J.\ Biomed.\ Mater.\ Res.\ Part\ A \textbf{102A},
  254--274 (2014).
\input{Kaur_jbmra_RevBioactiveGlass_2014}

\bibitem{Inoue_imr_ironBMG_2013}
C.~Suryanarayana and A.~Inoue, \emph{Iron-based bulk metallic glasses}, Int.\
  Mater.\ Rev. \textbf{58}, 131--166 (2013).
\input{Inoue_imr_ironBMG_2013}

\bibitem{Thouas_mser_metalImplant_2015}
Q.~Chen and G.~A. Thouas, \emph{Metallic implant biomaterials}, Mat.\ Sci.\
  Eng.\ R \textbf{87}, 1--57 (2015).
\input{Thouas_mser_metalImplant_2015}

\bibitem{Staiger_biomat_MgOrtho_2006}
M.~P. Staiger, A.~M. Pietak, J.~Huadmai, and G.~Dias, \emph{Magnesium and its
  alloys as orthopedic biomaterials: A review}, Biomaterials \textbf{27},
  1728--1734 (2006).
\input{Staiger_biomat_MgOrtho_2006}

\bibitem{Walker_jbmrb_MgBioPerspec_2014}
J.~Walker, S.~Shadanbaz, T.~B.~F. Woodfield, M.~P. Staiger, and G.~J. Dias,
  \emph{Magnesium biomaterials for orthopedic application: A review from a
  biological perspective}, J.\ Biomed.\ Mater.\ Res.\ Part\ B \textbf{102B},
  1316--1331 (2014).
\input{Walker_jbmrb_MgBioPerspec_2014}

\bibitem{Farraro_jbm_RevolOrtho_2014}
K.~F. Farraro, K.~E. Kim, S.~L.-Y. Woo, J.~R. Flowers, and M.~B. McCullough,
  \emph{Revolutionizing orthopaedic biomaterials: The potential of
  biodegradable and bioresorbable magnesium-based materials for functional
  tissue engineering}, J.\ Biomech. \textbf{47}, 1979--1986 (2014).
\input{Farraro_jbm_RevolOrtho_2014}

\bibitem{Chen_actabiomat_MgAdv_2014}
Y.~Chen, Z.~Xu, C.~Smith, and J.~Sankar, \emph{Recent advances on the
  development of magnesium alloys for biodegradable implants}, Acta\ Biomater.
  \textbf{10}, 4561--4573 (2014).
\input{Chen_actabiomat_MgAdv_2014}

\bibitem{Jafari_jom_MgSCC-CF_2015}
S.~Jafari, S.~E. Harandi, and R.~K.~S. Raman, \emph{A Review of
  Stress-Corrosion Cracking and Corrosion Fatigue of Magnesium Alloys for
  Biodegradable Implant Applications}, JOM \textbf{67}, 1143--1153 (2015).
\input{Jafari_jom_MgSCC-CF_2015}

\bibitem{Agarwal_msec_MgRevCBS_2016}
S.~Agarwal, J.~Curtin, B.~Duffy, and S.~Jaiswal, \emph{Biodegradable magnesium
  alloys for orthopaedic applications: A review on corrosion, biocompatibility
  and surface modifications}, Mat.\ Sci.\ Eng.\ C \textbf{68}, 948--963 (2016).
\input{Agarwal_msec_MgRevCBS_2016}

\bibitem{Narayanan_pms_MAO_2014}
T.~S.~N.~S. Narayanan, I.~S. Park, and M.~H. Lee, \emph{Strategies to improve
  the corrosion resistance of microarc oxidation (MAO) coated magnesium alloys
  for degradable implants: Prospects and challenges}, Prog.\ Mater.\ Sci.
  \textbf{60}, 1--71 (2014).
\input{Narayanan_pms_MAO_2014}

\bibitem{Kraus_actabiomat_Mginvivo_2012}
T.~Kraus, S.~F. Fischerauer, A.~C. H{\"a}nzi, P.~J. Uggowitzer, J.~F.
  L{\"o}ffler, and A.~M. Weinberg, \emph{Magnesium alloys for temporary
  implants in osteosynthesis: In vivo studies of their degradation and
  interaction with bone}, Acta\ Biomater. \textbf{8}, 1230--1238 (2012).
\input{Kraus_actabiomat_Mginvivo_2012}

\bibitem{Song_CS_BioDC_2007}
G.~Song, \emph{Control of biodegradation of biocompatable magnesium alloys},
  Corros.\ Sci. \textbf{49}, 1696--1701 (2007).
\input{Song_CS_BioDC_2007}

\bibitem{Ibrahim_msec_PostFab_2017}
H.~Ibrahim, S.~N. Esfahani, B.~Poorganji, D.~Dean, and M.~Elahinia,
  \emph{Resorbable bone fixation alloys, forming, and post-fabrication
  treatments}, Mat.\ Sci.\ Eng.\ C \textbf{70}, 870--888 (2017).
\input{Ibrahim_msec_PostFab_2017}

\bibitem{Cheng_pms_SPMG_2011}
Y.~Q. Cheng and E.~Ma, \emph{Atomic-level structure and structure-property
  relationship in metallic glasses}, Prog.\ Mater.\ Sci. \textbf{56}, 379--473
  (2011).
\input{Cheng_pms_SPMG_2011}

\bibitem{Scully_jmr_BMG-CMP_2007}
J.~R. Scully, A.~Gebert, and J.~H. Payer, \emph{Corrosion and related
  mechanical properties of bulk metallic glasses}, J.\ Mater.\ Res.
  \textbf{22}, 302--313 (2007).
\input{Scully_jmr_BMG-CMP_2007}

\bibitem{Zberg_NMAT_MgZnCa_2009}
B.~Zberg, P.~J. Uggowitzer, and J.~F. L{\"o}ffler, \emph{MgZnCa glasses without
  clinically observable hydrogen evolution for biodegradable implants}, Nat.\
  Mater. \textbf{8}, 887--891 (2009).
\input{Zberg_NMAT_MgZnCa_2009}

\bibitem{Lewandowski_phml_plasticMG_2005}
J.~J. Lewandowski, W.~H. Wang, and A.~L. Greer, \emph{Intrinsic plasticity or
  brittleness of metallic glasses}, Philos.\ Mag.\ Lett. \textbf{85}, 77--87
  (2005).
\input{Lewandowski_phml_plasticMG_2005}

\bibitem{Chen_armr_MG-duct_2008}
M.~Chen, \emph{Mechanical Behavior of Metallic Glasses: Microscopic
  Understanding of Strength and Ductility}, Annu.\ Rev.\ Mater.\ Res.
  \textbf{38}, 445--469 (2008).
\input{Chen_armr_MG-duct_2008}

\bibitem{Inoue_stabilization_2000}
A.~Inoue, \emph{Stabilization of metallic supercooled liquid and bulk amorphous
  alloys}, Acta\ Mater. \textbf{48}, 279--306 (2000).
\input{Inoue_stabilization_2000}

\bibitem{Li_acscombsci_numMG_2017}
Y.~Li, S.~Zhao, Y.~Liu, P.~Gong, and J.~Schroers, \emph{How Many Bulk Metallic
  Glasses Are There?}, ACS\ Comb.\ Sci. \textbf{19}, 687--693 (2017).
\input{Li_acscombsci_numMG_2017}

\bibitem{Ren_sciadv_MGexpML_2018}
F.~Ren, L.~Ward, T.~Williams, K.~J. Laws, C.~Wolverton, J.~Hattrick-Simpers,
  and A.~Mehta, \emph{Accelerated discovery of metallic glasses through
  iteration of machine learning and high-throughput experiments}, Sci.\ Adv.
  \textbf{4}, eaaq1566 (2018).
\input{Ren_sciadv_MGexpML_2018}

\bibitem{Vincent_jncs_thermoBMG_2011}
S.~Vincent, D.~R. Peshwe, B.~S. Murty, and J.~Bhatt, \emph{Thermodynamic
  prediction of bulk metallic glass forming alloys in ternary Zr-Cu-X (X = Ag,
  Al, Ti, Ga) systems}, J.\ Non-Cryst.\ Solids \textbf{357}, 3495--3499 (2011).
\input{Vincent_jncs_thermoBMG_2011}

\bibitem{Laws_ncomm_BMGmodel_2015}
K.~J. Laws, D.~B. Miracle, and M.~Ferry, \emph{A predictive structural model
  for bulk metallic glasses}, Nat.\ Commun. \textbf{6}, 8123 (2015).
\input{Laws_ncomm_BMGmodel_2015}

\bibitem{zhang2015origin}
K.~Zhang, B.~Dice, Y.~Liu, J.~Schroers, M.~D. Shattuck, and C.~S. O'{H}ern,
  \emph{On the origin of multi-component bulk metallic glasses: Atomic size
  mismatches and de-mixing}, J.\ Chem.\ Phys. \textbf{143}, 054501 (2015).
\input{zhang2015origin}

\bibitem{LL_jap_compGFA_2010}
D.~V. Louzguine-Luzgin, D.~B. Miracle, L.~Louzguina-Luzgina, and A.~Inoue,
  \emph{Comparative analysis of glass-formation in binary, ternary, and
  multicomponent alloys}, J.\ Appl.\ Phys. \textbf{108}, 103511 (2010).
\input{LL_jap_compGFA_2010}

\bibitem{greer1993confusion}
A.~L. Greer, \emph{Confusion by design}, Nature \textbf{366}, 303--304 (1993).
\input{greer1993confusion}

\bibitem{curtarolo:art112}
E.~Perim, D.~Lee, Y.~Liu, C.~Toher, P.~Gong, Y.~{Li}, W.~N. Simmons, O.~Levy,
  J.~J. Vlassak, J.~Schroers, and S.~Curtarolo, \emph{Spectral descriptors for
  bulk metallic glasses based on the thermodynamics of competing crystalline
  phases}, Nat.\ Commun. \textbf{7}, 12315 (2016).
\input{curtarolo:art112}

\bibitem{aflowlib}
W.~Setyawan and S.~Curtarolo, \emph{{AFLOWLIB}: Ab-initio Electronic Structure
  Library Database}, http://www.aflow.org (2011).
\input{aflowlib}

\bibitem{curtarolo:art92}
R.~H. Taylor, F.~Rose, C.~Toher, O.~Levy, K.~Yang, M.~{Buongiorno Nardelli},
  and S.~Curtarolo, \emph{A {REST}ful {API} for exchanging materials data in
  the {AFLOWLIB}.org consortium}, Comput.\ Mater.\ Sci. \textbf{93}, 178--192
  (2014).
\input{curtarolo:art92}

\bibitem{curtarolo:art75}
S.~Curtarolo, W.~Setyawan, S.~Wang, J.~Xue, K.~Yang, R.~H. Taylor, L.~J.
  Nelson, G.~L.~W. Hart, S.~Sanvito, M.~{Buongiorno Nardelli}, N.~Mingo, and
  O.~Levy, \emph{{AFLOWLIB.ORG}: A distributed materials properties repository
  from high-throughput {\it ab initio} calculations}, Comput.\ Mater.\ Sci.
  \textbf{58}, 227--235 (2012).
\input{curtarolo:art75}

\bibitem{villars:factors}
P.~Villars, \emph{Factors Governing Crystal Structures}, in \emph{Crystal
  Structures of Intermetallic Compounds}, edited by J.~H. Westbrook and R.~L.
  Fleisher (Wiley, New York, 2000), pp. 1--49.
\input{villars:factors}

\bibitem{Oses_CHULL_JCIM-pub_2018}
C.~Oses, E.~Gossett, D.~Hicks, F.~Rose, M.~J. Mehl, E.~Perim, I.~Takeuchi,
  S.~Sanvito, M.~Scheffler, Y.~Lederer, O.~Levy, C.~Toher, and S.~Curtarolo,
  \emph{{AFLOW-CHULL}: Cloud-Oriented Platform for Autonomous Phase Stability
  Analysis}, J.\ Chem.\ Inf.\ Model. \textbf{58}, 2477--2490 (2018).
\input{Oses_CHULL_JCIM-pub_2018}

\bibitem{aflow_fleet_chapter}
{C. Toher \textit{et al.}}, \emph{The {AFLOW} Fleet for Materials Discovery},
  in \emph{Handbook of Materials Modeling}, edited by W.~Andreoni and S.~Yip
  (Springer International Publishing, Cham, Switzerland, 2018), pp. 1--28,
  \doi{10.1007/978-3-319-42913-7_63-1}.
\input{aflow_fleet_chapter}

\bibitem{aflowPAPER18}
C.~Oses, C.~Toher, and S.~Curtarolo, \emph{Data-driven design of inorganic
  materials with the Automatic Flow Framework for Materials Discovery}, MRS\
  Bull. \textbf{43}, 670--675 (2018).
\input{aflowPAPER18}

\bibitem{curtarolo:art65}
S.~Curtarolo, W.~Setyawan, G.~L.~W. Hart, M.~Jahn\'{a}tek, R.~V. Chepulskii,
  R.~H. Taylor, S.~Wang, J.~Xue, K.~Yang, O.~Levy, M.~J. Mehl, H.~T. Stokes,
  D.~O. Demchenko, and D.~Morgan, \emph{{AFLOW}: An automatic framework for
  high-throughput materials discovery}, Comput.\ Mater.\ Sci. \textbf{58},
  218--226 (2012).
\input{curtarolo:art65}

\bibitem{curtarolo:art104}
C.~E. Calderon, J.~J. Plata, C.~Toher, C.~Oses, O.~Levy, M.~Fornari, A.~Natan,
  M.~J. Mehl, G.~L.~W. Hart, M.~{Buongiorno Nardelli}, and S.~Curtarolo,
  \emph{The {AFLOW} standard for high-throughput materials science
  calculations}, Comput.\ Mater.\ Sci. \textbf{108 Part A}, 233--238 (2015).
\input{curtarolo:art104}

\bibitem{kresse_vasp}
G.~Kresse and J.~Hafner, \emph{{\it Ab initio} molecular dynamics for liquid
  metals}, Phys.\ Rev.\ B \textbf{47}, 558--561 (1993).
\input{kresse_vasp}

\bibitem{vasp_prb1996}
G.~Kresse and J.~Furthm\"{u}ller, \emph{Efficient iterative schemes for {\it ab
  initio} total-energy calculations using a plane-wave basis set}, Phys.\ Rev.\
  B \textbf{54}, 11169--11186 (1996).
\input{vasp_prb1996}

\bibitem{PBE}
J.~P. Perdew, K.~Burke, and M.~Ernzerhof, \emph{Generalized Gradient
  Approximation Made Simple}, Phys.\ Rev.\ Lett. \textbf{77}, 3865--3868
  (1996).
\input{PBE}

\bibitem{PAW}
P.~E. Bl\"{o}chl, \emph{Projector augmented-wave method}, Phys.\ Rev.\ B
  \textbf{50}, 17953--17979 (1994).
\input{PAW}

\bibitem{kresse_vasp_paw}
G.~Kresse and D.~Joubert, \emph{From ultrasoft pseudopotentials to the
  projector augmented-wave method}, Phys.\ Rev.\ B \textbf{59}, 1758--1775
  (1999).
\input{kresse_vasp_paw}

\bibitem{aflowANRL}
M.~J. Mehl, D.~Hicks, C.~Toher, O.~Levy, R.~M. Hanson, G.~L.~W. Hart, and
  S.~Curtarolo, \emph{The {AFLOW} Library of Crystallographic Prototypes: Part
  1}, Comput.\ Mater.\ Sci. \textbf{136}, S1--S828 (2017).
\input{aflowANRL}

\bibitem{anrl_pt2_2018}
D.~Hicks, M.~J. Mehl, E.~Gossett, C.~Toher, O.~Levy, R.~M. Hanson, G.~L.~W.
  Hart, and S.~Curtarolo, \emph{The {AFLOW} Library of Crystallographic
  Prototypes: Part 2}, submitted arXiv:1806.07864  (2018).
\input{anrl_pt2_2018}

\bibitem{Guo_apl_CaAl-BMG_2004}
F.~Q. Guo, S.~J. Poon, and G.~J. Shiflet, \emph{CaAl-based bulk metallic
  glasses with high thermal stability}, Appl.\ Phys.\ Lett. \textbf{84}, 37--39
  (2004).
\input{Guo_apl_CaAl-BMG_2004}

\bibitem{inoue2004formation}
A.~Inoue and W.~Zhang, \emph{Formation, thermal stability and mechanical
  properties of Cu-Zr and Cu-Hf binary glassy alloy rods}, Mater.\ Trans.
  \textbf{45}, 584--587 (2004).
\input{inoue2004formation}

\bibitem{xia2006glass}
L.~Xia, D.~Ding, S.~T. Shan, and Y.~D. Dong, \emph{The glass forming ability of
  Cu-rich Cu-Hf binary alloys}, J.\ Phys.:\ Condens.\ Matter \textbf{18}, 3543
  (2006).
\input{xia2006glass}

\bibitem{wang2004bulk}
D.~Wang, Y.~{Li}, B.~B. Sun, M.~L. Sui, K.~Lu, and E.~Ma, \emph{Bulk metallic
  glass formation in the binary Cu-Zr system}, Appl.\ Phys.\ Lett. \textbf{84},
  4029--4031 (2004).
\input{wang2004bulk}

\bibitem{li2008matching}
Y.~{Li}, Q.~Guo, J.~A. Kalb, and C.~V. Thompson, \emph{Matching glass-forming
  ability with the density of the amorphous phase}, Science \textbf{322},
  1816--1819 (2008).
\input{li2008matching}

\bibitem{Xia_jap_NbNi-BMG_2006}
L.~Xia, W.~H. Li, S.~S. Fang, B.~C. Wei, and Y.~D. Dong, \emph{Binary Ni-Nb
  bulk metallic glasses}, J.\ Appl.\ Phys. \textbf{99}, 026103 (2006).
\input{Xia_jap_NbNi-BMG_2006}

\bibitem{leonhardt1999solidification}
M.~Leonhardt, W.~L\"{o}ser, and H.-G. Lindenkreuz, \emph{Solidification
  kinetics and phase formation of undercooled eutectic {Ni-Nb} melts}, Acta\
  Mater. \textbf{47}, 2961--2968 (1999).
\input{leonhardt1999solidification}

\bibitem{Chen_amet_PdSi-BMG_1969}
H.~S. Chen and D.~Turnbull, \emph{Formation, stability, and structure of
  palladium-silicon based alloy glasses}, Acta\ Metall. \textbf{17}, 1021--1031
  (1969).
\input{Chen_amet_PdSi-BMG_1969}

\bibitem{Wang_actamat_AlCuZr_2005}
D.~Wang, H.~Tan, and Y.~Li, \emph{Multiple maxima of GFA in three adjacent
  eutectics in Zr-Cu-Al alloy system - A metallographic way to pinpoint the
  best glass forming alloys}, Acta\ Mater. \textbf{53}, 2969--2979 (2005).
\input{Wang_actamat_AlCuZr_2005}

\bibitem{Kim_msea_AlCuZr-1_2006}
Y.~C. Kim, J.~C. Lee, P.~R. Cha, J.~P. Ahn, and E.~Fleury, \emph{Enhanced glass
  forming ability and mechanical properties of new Cu-based bulk metallic
  glasses}, Mat.\ Sci.\ Eng.\ A \textbf{437}, 248--253 (2006).
\input{Kim_msea_AlCuZr-1_2006}

\bibitem{Xu_prl_AlCuZr-2_2004}
D.~H. Xu, G.~Duan, and W.~L. Johnson, \emph{Unusual Glass-Forming Ability of
  Bulk Amorphous Alloys Based on Ordinary Metal Copper}, Phys.\ Rev.\ Lett.
  \textbf{92}, 245504 (2004).
\input{Xu_prl_AlCuZr-2_2004}

\bibitem{Inoue_mattrans_AlCuZr-3_1995}
A.~Inoue, \emph{High Strength Bulk Amorphous Alloys with Low Critical Cooling
  Rates}, Mater.\ Trans. \textbf{36}, 866--875 (1995).
\input{Inoue_mattrans_AlCuZr-3_1995}

\bibitem{Li_jac_AlNiZr-2_2012}
Y.~H. Li, W.~Zhang, C.~Dong, J.~B. Qiang, G.~Q. Xie, K.~Fujita, and A.~Inoue,
  \emph{Glass-forming ability and corrosion resistance of Zr-based Zr-Ni-Al
  bulk metallic glasses}, J.\ Alloys\ Compd. \textbf{536S}, S117--S121 (2012).
\input{Li_jac_AlNiZr-2_2012}

\bibitem{Senkov_jac_CaMgCu-BMG_2006}
O.~N. Senkov, J.~M. Scott, and D.~B. Miracle, \emph{Composition range and glass
  forming ability of ternary Ca-Mg-Cu bulk metallic glasses}, J.\ Alloys\
  Compd. \textbf{424}, 394--399 (2006).
\input{Senkov_jac_CaMgCu-BMG_2006}

\bibitem{Laws_jac_AgCaCuMg_2012}
K.~J. Laws, K.~F. Shamlaye, J.~D. Cao, J.~P. Scicluna, and M.~Ferry,
  \emph{Locating new Mg-based bulk metallic glasses free of rare earth
  elements}, J.\ Alloys\ Compd. \textbf{542}, 105--110 (2012).
\input{Laws_jac_AgCaCuMg_2012}

\bibitem{Gu_jmr_MgCaZn_2005}
X.~Gu, G.~J. Shiflet, F.~Q. Guo, and S.~J. Poon, \emph{Mg-Ca-Zn bulk metallic
  glasses with high strength and significant ductility}, J.\ Mater.\ Res.
  \textbf{20}, 1935--1938 (2005).
\input{Gu_jmr_MgCaZn_2005}

\bibitem{Senkov_jncs_CaMgZn-BMG_2005}
O.~N. Senkov and J.~M. Scott, \emph{Glass forming ability and thermal stability
  of ternary Ca-Mg-Zn bulk metallic glasses}, J.\ Non-Cryst.\ Solids
  \textbf{351}, 3087--3094 (2005).
\input{Senkov_jncs_CaMgZn-BMG_2005}

\bibitem{Park_jmr_CaMgZn-2_2004}
E.~S. Park and D.~H. Kim, \emph{Formation of Ca-Mg-Zn bulk glassy alloy by
  casting into cone-shaped copper mold}, J.\ Mater.\ Res. \textbf{19}, 685--688
  (2004).
\input{Park_jmr_CaMgZn-2_2004}

\bibitem{Ma_jmr_CuMgY-1_2005}
H.~Ma, Q.~Zheng, J.~Xu, Y.~Li, and E.~Ma, \emph{Doubling the critical size for
  bulk metallic glass formation in the Mg-Cu-Y ternary system}, J.\ Mater.\
  Res. \textbf{20}, 2252--2255 (2005).
\input{Ma_jmr_CuMgY-1_2005}

\bibitem{Inoue_mattrans_MgCuY_1991}
A.~Inoue, A.~Kato, T.~Zhang, S.~G. Kim, and T.~Masumoto, \emph{Mg-Cu-Y
  Amorphous Alloys with High Mechanical Strengths Produced by a Metallic Mold
  Casting Method}, Mater.\ Trans. \textbf{32}, 609--616 (1991).
\input{Inoue_mattrans_MgCuY_1991}

\bibitem{Inoue_mattrans_AlNiZr-1_1990}
A.~Inoue, T.~Zhang, and T.~Masumoto, \emph{Zr-Al-Ni Amorphous Alloys with High
  Glass Transition Temperature and Significant Supercooled Liquid Region},
  Mater.\ Trans. \textbf{31}, 177--183 (1990).
\input{Inoue_mattrans_AlNiZr-1_1990}

\bibitem{Wang_pms_elasticMG_2012}
W.~H. Wang, \emph{The elastic properties, elastic models and elastic
  perspectives of metallic glasses}, Prog.\ Mater.\ Sci. \textbf{57}, 487--656
  (2012).
\input{Wang_pms_elasticMG_2012}

\bibitem{Zhang_jac_elastic-correlations_2007}
Y.~Zhang and A.~L. Greer, \emph{Correlations for predicting plasticity or
  brittleness of metallic glasses}, J.\ Alloys\ Compd. \textbf{434-435}, 2--5
  (2007).
\input{Zhang_jac_elastic-correlations_2007}

\bibitem{webelements}
{W}eb{E}lements, https://www.webelements.com/.
\input{webelements}

\bibitem{Vasconcelos_biomat_twoface_2016}
D.~M. Vasconcelos, S.~G. Santos, M.~Lamghari, and M.~A. Barbosa, \emph{The two
  faces of metal ions: From implants rejection to tissue repair/regeneration},
  Biomaterials \textbf{84}, 262--275 (2016).
\input{Vasconcelos_biomat_twoface_2016}

\bibitem{aflowsym}
D.~Hicks, C.~Oses, E.~Gossett, G.~Gomez, R.~H. Taylor, C.~Toher, M.~J. Mehl,
  O.~Levy, and S.~Curtarolo, \emph{{\it AFLOW-SYM}: platform for the complete,
  automatic and self-consistent symmetry analysis of crystals}, Acta\
  Crystallogr.\ Sect.\ A \textbf{74}, 184--203 (2018).
\input{aflowsym}

\bibitem{aflow_compare-updated_2018}
D.~Hicks, C.~Toher, D.~C. Ford, F.~Rose, C.~{De Santo}, O.~Levy, M.~J. Mehl,
  and S.~Curtarolo, \emph{{AFLOW-XTAL-MATCH}: Automated method for quantifying
  the structural similarity of materials and identifying unique crystal
  prototypes}, in preparation  (2019).
\input{aflow_compare-updated_2018}

\bibitem{Amand_smet_CaSr-based-MG_1978}
R.~S. Amand and B.~C. Giessen, \emph{Easy glass formation in simple metal
  alloys: amorphous metals containing calcium and strontium}, Scr.\ Metall.
  \textbf{12}, 1021--1026 (1978).
\input{Amand_smet_CaSr-based-MG_1978}

\bibitem{Kim_mattrans_CuMg-MG_1990}
S.~G. Kim, A.~Inoue, and T.~Masumoto, \emph{High Mechanical Strengths of
  Mg-Ni-Y and Mg-Cu-Y Amorphous Alloys with Significant Supercooled Liquid
  Region}, Mater.\ Trans. \textbf{31}, 929--934 (1990).
\input{Kim_mattrans_CuMg-MG_1990}

\bibitem{Calka_smet_MgZn-MG_1977}
A.~Calka, M.~Madhava, D.~E. Polk, B.~C. Giessen, H.~Matyja, and J.~{Vander
  Sande}, \emph{A transition-metal-free amorphous alloy: Mg$_{.70}$Zn$_{.30}$},
  Scr.\ Metall. \textbf{11}, 65--70 (1977).
\input{Calka_smet_MgZn-MG_1977}

\bibitem{Guo_scrmat_Mg-duct_2007}
F.~Guo, S.~J. Poon, X.~Gu, and G.~J. Shiflet, \emph{Low-density Mg-rich
  metallic glasses with bending ductility}, Scr.\ Mater. \textbf{56}, 689--692
  (2007).
\input{Guo_scrmat_Mg-duct_2007}

\bibitem{Widom_prb_CaMG-elastic_2011}
M.~Widom, B.~Sauerwine, A.~M. Cheung, S.~J. Poon, P.~Tong, D.~Louca, and G.~J.
  Shiflet, \emph{Elastic properties of Ca-based metallic glasses predicted by
  first-principles simulations}, Phys.\ Rev.\ B \textbf{84}, 054206 (2011).
\input{Widom_prb_CaMG-elastic_2011}

\bibitem{Laws_actamat_MgDuct_2016}
K.~J. Laws, D.~Granata, and J.~F. L{\"o}ffler, \emph{Alloy design strategies
  for sustained ductility in Mg-based amorphous alloys - Tackling structural
  relaxation}, Acta\ Mater. \textbf{103}, 735--745 (2016).
\input{Laws_actamat_MgDuct_2016}

\bibitem{Park_jncs_AgCuMgY_2001}
E.~S. Park, H.~G. Kang, W.~T. Kim, and D.~H. Kim, \emph{The effect of Ag
  addition on the glass-forming ability of Mg-Cu-Y metallic glass alloys}, J.\
  Non-Cryst.\ Solids \textbf{279}, 154--160 (2001).
\input{Park_jncs_AgCuMgY_2001}

\bibitem{Li_jncs_AgCaMgZn_2015}
H.~Li, S.~Pang, Y.~Liu, P.~K. Liaw, and T.~Zhang, \emph{In vitro investigation
  of Mg-Zn-Ca-Ag bulk metallic glasses for biomedical applications}, J.\
  Non-Cryst.\ Solids \textbf{427}, 134--138 (2015).
\input{Li_jncs_AgCaMgZn_2015}

\bibitem{Amiya_mattrans_AgCaCuMg_2002}
K.~Amiya and A.~Inoue, \emph{Formation and Thermal Stability of Ca-Mg-Ag-Cu
  Bulk Glassy Alloys}, Mater.\ Trans. \textbf{43}, 2578--2581 (2002).
\input{Amiya_mattrans_AgCaCuMg_2002}

\bibitem{curtarolo:art128}
F.~Rose, C.~Toher, E.~Gossett, C.~Oses, M.~{Buongiorno Nardelli}, M.~Fornari,
  and S.~Curtarolo, \emph{{AFLUX}: The {LUX} materials search {API} for the
  {AFLOW} data repositories}, Comput.\ Mater.\ Sci. \textbf{137}, 362--370
  (2017).
\input{curtarolo:art128}

\end{thebibliography}
\end{document}